\documentclass[default,times]{aastex701}

\usepackage{amsmath}

\received{}
\revised{}
\accepted{}
\submitjournal{ApJL}

\begin{document}

\title{A Model for the Enhanced Production Rate of Early-Type Hypervelocity Stars 
  in the Galactic Halo
}

\correspondingauthor{F.K. Liu}
\email{fkliu@pku.edu.cn}

\author[0000-0002-4211-9523]{Chunyang Cao}
\affiliation{Department of Astronomy, School of Physics, Peking University, 
 Beijing 100871, People's Republic of China}
\email{icevoicey@pku.edu.cn}

\author[0000-0002-5310-3084]{F.K. Liu}
\affiliation{Department of Astronomy, School of Physics, Peking University, 
Beijing 100871, People's Republic of China}
\affiliation{Kavli Institute for Astronomy and Astrophysics, Peking University, 
 Beijing 100871, People's Republic of China}
\email{fkliu@pku.edu.cn}

\author[0000-0003-3950-9317]{Xian Chen}
\affiliation{Department of Astronomy, School of Physics, Peking University, 
 Beijing 100871, People's Republic of China}
\affiliation{Kavli Institute for Astronomy and Astrophysics, Peking University, 
 Beijing 100871, People's Republic of China}
\email{xian.chen@pku.edu.cn}

\author[0000-0001-6530-0424]{Shuo Li}
\affiliation{National Astronomical Observatories, Chinese Academy of Sciences, 
Beijing 100012, People's Republic of China}
\email{lishuo@nao.cas.cn}

\begin{abstract}
About twenty late B-type hypervelocity stars (HVSs) traveling faster than 
the Galactic escape velocity have been discovered in the Galactic halo, 
many of which were ejected from the Galactic center (GC). 
Recently, we have advocated that these HVSs most likely
formed in the nuclear star cluster (NSC) $150$--$500\, \rm{Myr}$ ago and 
were predominantly ejected via the gravitational slingshot of 
a past intermediate-mass black hole (IMBH) orbiting 
the supermassive black hole (SMBH) Sgr~A$^{*}$.
Here we explore the constraints of the production rate of young HVSs 
on the star formation region of the NSC.  We propose that
the young HVS progenitors are born in a lopsided eccentric disk 
that is comparable in radius to the NSC. 
By numerically tracking the orbital evolution of disk stars, 
we find that they undergo rapid angular momentum relaxation 
at formation due to eccentric disk instability, 
and that their slingshot interactions 
with the SMBH-IMBH binary at distances $\simeq 100\, \rm{au}$
% $\leq 50$--$400\, \rm{au}$ 
produce HVSs at a rate of $10^{-5}$--$10^{-4}\, \rm{yr}^{-1}$. 
The rate is expected to trace the disk formation history, 
increasing with the accumulation of disk stars and dropping rapidly 
after the star formation stopped at $150\, \rm{Myr}$ ago. 
The rate is consistent with the observation and 
orders of magnitude higher than 
that expected for an old relaxed population in the literature, enhanced
due to the gravitational torque from the non-spherical GC potential and 
radial velocity anisotropies of the disk stars. 
Our results imply that young HVSs 
should have a distinct radial and angular distribution from old ones. 
\end{abstract}

\section{Introduction} \label{sec:introduction}

The existence of a supermassive black hole (SMBH) Sgr~A$^{*}$ 
at the Galactic center (GC) has been revealed with the stellar motions, 
as well as its proper motion and direct image by the Event Horizon Telescope 
\citep[EHT;][]{eckart_observations_1996,reid_proper_2020,collaboration_first_2022}. 
The SMBH is also manifested by the discovery of GC-origin hypervelocity stars 
\citep[HVSs;][]{brown_discovery_2005}, 
because their extreme velocities (exceeding the Galactic escape velocity)
can only be obtained through close encounters with an SMBH,  
either via the Hills mechanism, 
in which a stellar binary is tidally separated by the SMBH and 
one of the components is energetically ejected \citep{hills_hypervelocity_1988}, 
or via gravitational slingshot interactions with an SMBH binary \citep{yu_ejection_2003}. 
% \tg{
% So far, more than twenty HVSs have been identified 
% \citep[][]{brown_discovery_2005,brown_mmt_2014,koposov_discovery_2020,deng_old_2026}, 
% most of which are young (of late B-type with mass 2.5--4$M_{\odot}$) 
% and discovered in the Galactic halo by the MMT HVS survey \citep[][]{brown_mmt_2014}. 
% With the proper motions measured by Gaia \citep{2021A&A...649A...1G}, 
% about one third of them are found to be ejected from the GC 
% \citep[][]{brown_gaia_2018,irrgang_hypervelocity_2018,kreuzer_hypervelocity_2020,
% irrgang_blue_2021,han_hypervelocity_2025} 
% over the past 50--250$\, \rm{Myr}$ \citep[][]{brown_mmt_2014}. 
% }
So far, over twenty HVSs have been identified 
\citep[e.g.,][]{brown_mmt_2014,koposov_discovery_2020,deng_old_2026}, 
most of which are of late B-type with mass 2.5--4$M_{\odot}$ 
and discovered in the Galactic halo by the MMT survey \citep[][]{brown_mmt_2014}. 
With Gaia \citep{2021A&A...649A...1G} proper motions, 
% with proper motions measured mainly by Gaia \citep{2021A&A...649A...1G}, 
about one third of them are found to trace back to the GC 
% be ejected from the GC 
% have trajectories tracing 
\citep[][]{brown_gaia_2018,irrgang_hypervelocity_2018,kreuzer_hypervelocity_2020,
irrgang_blue_2021,han_hypervelocity_2025},
revealing a continuous ejection of young HVSs from the GC over the past 50--250$\, \rm{Myr}$ 
\citep[][]{brown_mmt_2014}.

% The GC-origin HVSs are powerful probes of the dynamical environment of the GC. 
% Particularly, they 
% an intermediate-mass black hole (IMBH) companion of the SMBH Sgr~A$^{*}$ 
% \citep[e.g.,][]{yu_ejection_2003}. 
% A long-standing puzzle is the unexpected cutoff at $\simeq 700\, \rm km\, s^{-1}$ 
% in the velocity distribution of 
% The velocity distribution of the late B-type HVSs 
% exhibits an unexpected cutoff at \sim 700\, \mathrm{km\, s^{-1}}, 
% which remains a long-standing puzzle 
% as about half of them 
% \citep[e.g.,][]{sesana_hypervelocity_2007,zhang_spatial_2010,
% rossi_velocity_2014,generozov_constraints_2022}. 
% model the 
% diffusion and interaction of the 500 Myr–population with an SMBH-IMBH binary in a Monte Carlo approach

It has been a long-standing puzzle that the velocity distribution of the late B-type HVSs 
shows an unexpected cutoff at $\simeq 700\, \rm km\, s^{-1}$ 
\citep[e.g.,][]{sesana_hypervelocity_2007,zhang_spatial_2010,
rossi_velocity_2014,generozov_constraints_2022}. 
In a recent work \citep{cao_recent_2025}, 
we showed that the velocity cutoff can be resolved by considering 
an intermediate-mass black hole (IMBH) around Sgr~A$^{*}$ 
kicking away slowly approaching compact stellar binaries 
and effectively inhibiting their tidal separations. 
And by Monte Carlo modeling the velocity and Galactocentric distance distribution 
of late B-type HVSs, 
we inferred that the IMBH should have a mass of $\simeq 1.5\times 10^{4} M_{\odot}$ 
and merged with Sgr~A$^{*}$ $\sim 10\, \rm{Myr}$ ago. 
% by exploiting the GC-origin HVSs in the halo, 
% especially the puzzling cutoff at $\simeq 700\, \rm km\, s^{-1}$ 
% in their velocity distribution
% \citep[e.g.,][]{sesana_hypervelocity_2007,zhang_spatial_2010,
% rossi_velocity_2014,generozov_constraints_2022}, 
% we have proposed in \cite{cao_recent_2025} that the GC should have hosted 
% a $\simeq 1.5\times 10^{4} M_{\odot}$ intermediate-mass black hole (IMBH) 
% orbiting and merging with the SMBH Sgr~A$^{*}$ $\sim 10\, \rm{Myr}$ ago. 
% we have exploited the GC-origin HVSs in the halo and 
% And through Monte Carlo simulations of 
% by numerically modeling %  in a Monte Carlo approach, 
% the stellar diffusion and interaction with the SMBH-IMBH binary, 
The SMBH-IMBH binary scenario also provides a unified explanation for 
% reproduces the velocity and Galactocentric distance distribution of late B-type HVSs, 
the age of S-stars \citep[$\lesssim 15\, \rm{Myr}$;][]{habibi_twelve_2017} 
and the proper motion of Sgr~A$^{*}$ \citep{reid_proper_2020}. 
Our finding is to some extent in tension with \cite{evans_constraints_2023}, 
who claimed that a $\geq 500M_{\odot}$ IMBH merged with Sgr~A$^{*}$ 
within the past $\sim 10\, \rm{Myr}$ can be excluded 
based on the null detection of new HVS in Gaia DR3 \citep{2021A&A...649A...1G}. 
However, their conclusion is obtained based on the assumption 
that the interaction loss cone of stars and the SMBH-IMBH binary 
is always efficiently refilled (full loss cone regime),
% supplied to interact with the SMBH-IMBH binary 
which has not been justified and 
may not be valid for the SMBH-IMBH binary with 
an orbit of hundreds of astronomical units in the GC. 

The late B-type HVSs, based on their main-sequence lifespans and flight times, 
most likely originate in the stellar population of the nuclear star cluster (NSC) 
formed over the past $150$--$500\, \rm{Myr}$ 
\citep[hereafter $500\, \rm{Myr}$--population;][]
{nishiyama_spectroscopically_2016,schodel_milky_2020,gallego-cano_age_2026}. 
However, according to the observational production rate 
of late B-type HVSs ($10^{-7}$--$10^{-6}\, {\rm yr^{-1}}$), 
the $500\, \rm{Myr}$--population are suggested to produce HVSs 
of all spectral types at a rate of $10^{-5}$--$10^{-4}\, {\rm yr^{-1}}$ 
\citep{brown_mmt_2014,brown_gaia_2018,cao_recent_2025}, 
which is several orders of magnitude higher than 
the theoretical expectation either with the Hills mechanism or 
the gravitational slingshot mechanism \citep{yu_ejection_2003} 
after scaling for the $\sim 1\%$--$4\%$ mass fraction of the $500\, \rm{Myr}$--population 
in the NSC \citep{schodel_milky_2020,gallego-cano_age_2026}. 
The rate could be enhanced significantly due to 
boosted dynamical relaxation from 
massive perturbers like giant molecular clouds \citep[GMCs;][]{perets_massive_2007}. 
% which possibly responsible for the recently reported old HVS candidate \citep{deng_old_2026}. 
Nevertheless, GMCs are generally confined to the distant nuclear stellar disk (NSD) 
and have little impact on the $500\, \rm{Myr}$--population of the NSC 
\citep[e.g.,][]{penoyre_disruptions_2025}. 
% ($r\gtrsim 10\, \rm{pc}$) 
% by considering young stars of the nuclear star cluster (NSC) 
% diffusing towards and interacting with a SMBH-IMBH binary 
% of mass ratio $\sim 5\times 10^{-3}$ that merged $\sim 10\, \rm{Myr}$ ago. 
% the young GC-origin HVSs in the halo were predominantly slingshot by a SMBH-IMBH binary. 
% The main-sequence lifespans of the late B-type HVSs suggest that 
% they must originate in a population of young stars of age $\sim 100\, \rm{Myr}$. 
% The best candidate is the stellar population 
% in the nuclear star cluster (NSC) formed over the past $150$--$500\, \rm{Myr}$ 
% \citep[hereafter $500\, \rm{Myr}$--population;][]
% {nishiyama_spectroscopically_2016,schodel_milky_2020,gallego-cano_age_2026}. 

In this Letter, we propose that the $500\, \rm{Myr}$--population of the NSC 
form in an eccentric stellar disk. 
We show that the enhancement of the HVS production rate is due both to 
the rapid angular momentum relaxation of the newly born stars 
because of the eccentric disk instability, 
and to the gravitational torque 
from the non-spheric GC potential and the radial velocity anisotropy of the disk stars.
In Section~\ref{sec:model}, we introduce our disk model by describing 
its formation and evolution in the GC environment. 
In Section~\ref{sec:rate}, we compute the HVS production rate for the disk stars. 
We summarize our findings and make a brief discussion in Section~\ref{sec:conclusions}.

\section{Eccentric Stellar Disk Model} \label{sec:model}

\subsection{Formation of the Stellar Disk} \label{subsec:formation}

We consider a scenario where a population of stars 
form in a gas stream funneling into the GC. 
The newborn stars inherit angular momentum from the gas stream, 
so their orbits have the same eccentricities $e_{\rm{d}}$ and aligned apsis, 
assembling a lopsided eccentric stellar disk. 
The disk is assumed to extend from $r_{\rm{in}} = 1.5\, \rm{pc}$ 
by referring to the inner cavity of the circum-nuclear disk 
\citep[CND;][]{becklin_farinfrared_1982,christopher_hcn_2005} 
to $r_{\rm{out}} \approx 20\, \rm{pc}$, 
about the external radius of the NSC \citep[e.g.,][]{chatzopoulos_old_2015}. 
The surface density is $\Sigma(r)\propto (rv_{\rm{ff}})^{-1}$ 
with $v_{\rm{ff}}$ the free-fall velocity. 
During the star formation over lookback time ($t_{\rm{lb}}$) 
$150$--$500\, \rm{Myr}$, 
the disk mass accumulates as 
\begin{eqnarray} \label{equ:disk_formation}
  M_{\rm{d}}(t_{\rm{lb}})= 
  \alpha_{\rm{sf}} \int^{500\, \rm{Myr}}_{\max({t_{\rm{lb}},150\, \rm{Myr}})}
  \mathrm{d}t,
\end{eqnarray} 
where we assume a constant star formation rate 
calibrated to $\alpha_{\rm{sf}}\approx 3.3\times 10^{-3}M_{\odot}\ \rm{yr}^{-1}$ 
such that at $0.5\leq r \leq 0.8\, \rm{pc}$ ($r \approx 3.5\, \rm{pc}$), 
the disk constitutes $4\%$ ($1\%$) of the NSC mass 
\citep[][]{schodel_milky_2020,gallego-cano_age_2026}.

\subsection{Early Evolution of the Stellar Disk} \label{subsec:early}

Given its small mass fraction in the NSC, 
the stellar disk is not a self-gravitational system, 
but evolve in the potential field of the GC. 
The stellar orbits can be approximated as a series of ellipses around the center SMBH. 
In a radial orbit period $T_{\rm{d}}$, 
their apsidal angles ($\omega$) precess by 
\citep[][]{binney_galactic_2011}
\begin{eqnarray} \label{equ:precess}
  \Delta \omega = 
  2L\int_{r_{1}}^{r_{2}}\frac{\rm{d}r}{r^{2}\sqrt{2[E-\Phi(r)]-L^{2}/r^{2}}}
  -2\pi
\end{eqnarray}
due to the extended mass distribution, 
where $r_{1}$ and $r_{2}$ are the pericenter and apocenter distances, 
$E$ and $L$ are the orbital energy and angular momentum per unit mass, 
and $\Phi$ is the GC potential. 
For the modeled disk, the precession timescale is quite short. 
For example, at the influence radius of Sgr~$\rm{A}^{*}$: $r_{\rm{i}}\approx 3\, \rm{pc}$, 
the typical precession angle is $\Delta \omega\approx -\pi/2$ 
(the minus sign indicates the retrograde precession direction) 
and the precession timescale is 
$T_{\rm{pre}}= |(2\pi/\Delta\omega) T_{\rm{d}}|\approx 0.5\, \rm{Myr}$. 
Therefore, any coherent processes, e.g., 
the resonant relaxation \citep{rauch_resonant_1996} 
and the Lidov-Kozai oscillation \citep{lidov_evolution_1962,kozai_secular_1962}, 
are suppressed given their much longer timescales 
\citep[e.g.,][]{merritt_orbits_2011,haas_rich_2016} and neglected in this study.

According to Equation~(\ref{equ:precess}), 
the disk stars with different $E$ and $L$ would 
precess differentially and gradually lose alignments. 
Meanwhile, their angular momentum evolve rapidly 
due to the so-called eccentric disk instability 
\citep{madigan_new_2009,gualandris_eccentric_2012}. 
Specifically, stars with smaller (larger) $L$ 
would precess behind (ahead) and feel torques from the bulk of the disk, 
which would further decrease (increase) their $L$. 
Because the instability relies heavily on the disk alignment, 
the rapid $L$-evolution sustains only for a few $T_{\rm{pre}}$ 
at the very beginning of the disk formation, 
after which the disk evolves from a lopsided spindle to an axisymmetric ring.

For illustration purpose, 
we simulate the early evolution of the stellar disk for $5\, \rm{Myr}$ in the GC. 
We use the N-body code \texttt{REBOUND} and 
its builtin IAS15 integrator \citep{rein_rebound_2012,rein_ias15_2015} 
to account for the internal gravity of disk stars. 
The GC potential is implemented analytically, 
modeled following \cite{penoyre_disruptions_2025} as a combination of: 
(i) a Kepler potential from the SMBH of mass 
$M_{\rm{SMBH}}= 4\times 10^{6}M_{\odot}$; 
(ii) a slightly axisymmetric (axis ratio 0.73) potential from the 
NSC of total stellar mass $6.1\times 10^{7}M_{\odot}$ \citep{chatzopoulos_old_2015}; 
and (iii) a significantly axisymmetric potential (axis ratio 0.37) from 
the nuclear stellar disk 
of total stellar mass $9.7\times 10^{8}M_{\odot}$ \citep{sormani_selfconsistent_2022}. 
The stellar disk is represented by 1,000 equal-mass particles 
with total mass $0.1M_{\rm{SMBH}}$, 
initially distributed around the SMBH 
following the density profile introduced in Section~\ref{subsec:formation}. 
We explore four initial disk configurations: 
\begin{enumerate}
  \item Run (A, B, C): the disk stars are aligned 
  with the same ascending node ($\Omega$) and azimuthal angle ($\phi$) and 
  have eccentricities: $e_{\rm{d}}=0.3,\, 0.5,\, 0.7$ 
  \footnote{
  The eccentricity is defined as $e_{\rm{d}}\equiv (r_{2}-r_{1})/(r_{1}+r_{2})$. 
  }; 
  \item Run (D): the disk stars have $e_{\rm{d}}=0.7$ and are non-aligned 
  with $\Omega$ and $\phi$ uniformly distributed in $[0,2\pi)$. 
\end{enumerate}
The inclinations $\theta$ are sampled from a Rayleigh distribution 
with $\langle \theta^{2}\rangle^{1/2} = 0.05$ \citep[e.g.,][]{lissauer_growth_1993}, 
and the mean anomalies are sampled uniformly in $[0,2\pi)$. 
For each disk configuration, 
we perform 20 independent realizations of the simulation 
to suppress statistical noise.

\begin{figure*}
  \centering
  \includegraphics[width=0.8\textwidth]{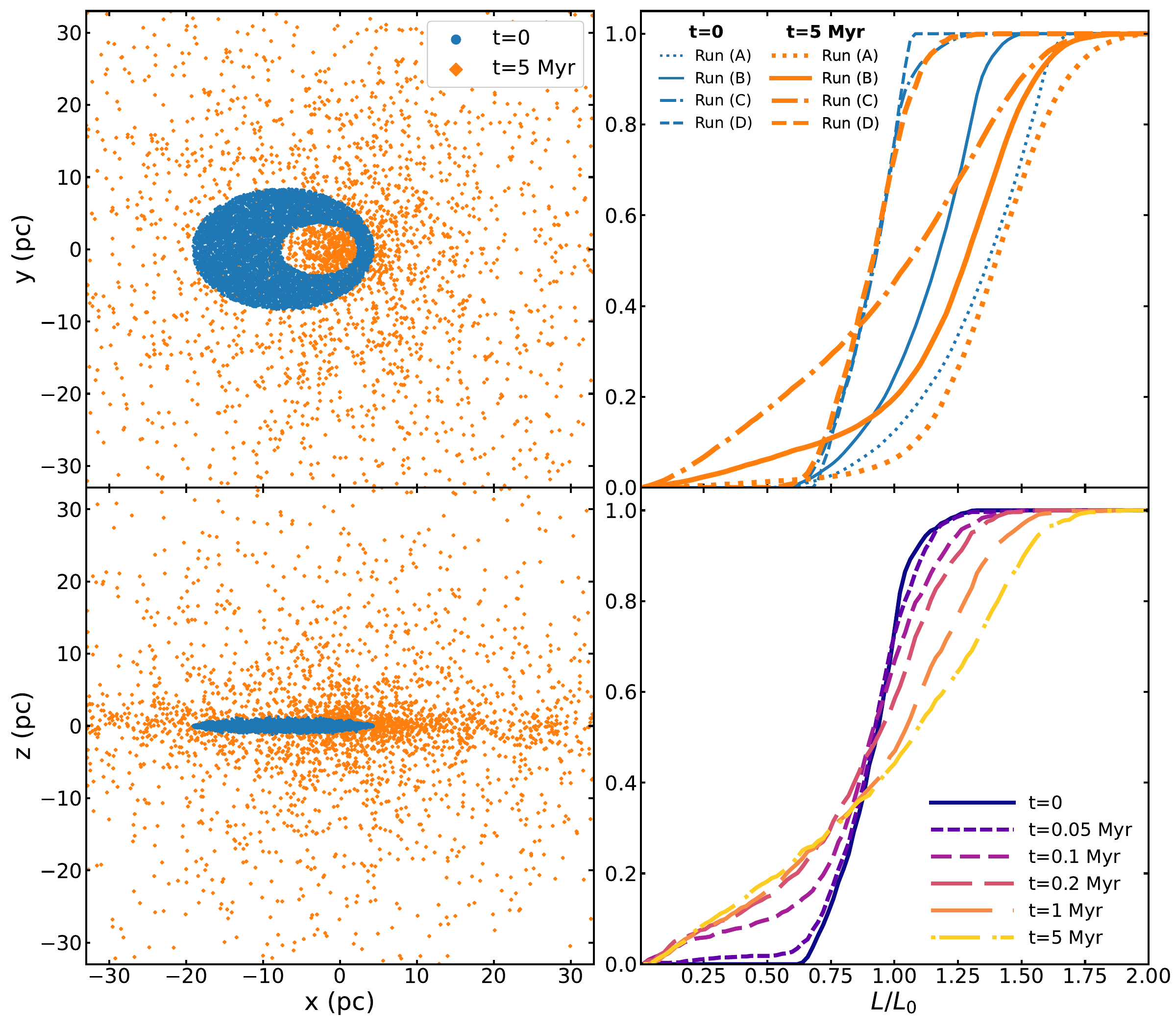}
  \caption{
  Left column: Initial (blue) and final (orange) spatial distribution 
  of the disk stars in Run (C) projected onto the x-y and x-z planes. 
  Right column: Cumulative distribution % function ($F$)
  of the scalar angular momentum ($L$) of disk stars normalized to 
  an arbitrary unit: $L_{\rm{0}}=300\, \rm{pc}\, \rm{km\, s^{-1}}$. 
  The upper panel shows the initial (thin blue) and final (thick orange) 
  distribution in different runs. 
  The lower panel shows the distribution snapped at different times in Run (C). 
  } 
  \label{fig:rebound}
\end{figure*}

In the left column of Figure \ref{fig:rebound}, 
we display the spatial distribution of the disk stars 
at the beginning and ending of Run (C). 
The stars initially have common apsides and quickly spread to 
an uniform apsis distribution (Kolmogorov-Smirnov test $p$--value$>0.05$) 
after about $1.8\, \rm{Myr}$ due to differential precession. 
Meanwhile, many stars emerge from the disk 
as their orbits become more inclined over time. 
In the upper right panel, 
we show the initial and final cumulative distribution function ($F$) of $L$ 
for the disk stars in different runs. 
The $L$-evolution is dominated by the eccentric disk instability 
rather than two-body relaxation, 
and thus is more effective for highly eccentric ($e_{\rm{d}}\gtrsim 0.4$) 
and aligned disks in Run (B) and (C) compared to that in Run (A) and (D). 
Likewise, the $L$-evolution in an initially aligned disk 
gradually slows down as the disk stars lose alignments, 
which is shown in the lower right panel. 
We note that limited by the computation efficiency, 
the simulation use a much smaller number of particles for the disk, 
which is expected to enlarge the two-body relaxation effect. 
Nevertheless, our results are not expected to be impacted 
given the minor role of two-body relaxation. 
We perform an additional run where the disk is represented by 4,000 particles 
and find no noticeable difference in the final $L$-distribution 
(two-sample Anderson-Darling test $p$--value$\approx 0.32$).

\subsection{Secular Evolution of the Stellar Disk} \label{subsec:secular}

After the early rapid evolution, 
the stellar disk settles into a quasi-steady stage and 
progressively relax due to two-body scattering. 
The relaxation rate can be denoted as \citep{chandrasekhar_dynamical_1943}: 
\begin{eqnarray} \label{equ:deflect}
  T_{\rm{r}}^{-1}=
  \frac{1}{L_{\rm{c}}^{2}}\frac{\mathrm{d}L^{2}}{\mathrm{d}t}=
  \frac{1}{v_{\rm{c}}^{2}} \frac{\mathrm{d}v^{2}}{\mathrm{d}t}\approx 
  \frac{8\pi G^{2}M\mu}{v_{\rm{c}}^{3}}\ln{\Lambda}, 
\end{eqnarray}
where $T_{\rm{r}}$ is the two-body relaxation timescale, 
$L_{\rm{c}}$ and $v_{\rm{c}}$ is the angular momentum and velocity of an circular orbit, 
$M$ is the total stellar mass, 
and $\mu \equiv \langle m^2 \rangle/\langle m \rangle$ 
is the ratio between the mean square mass and the mean mass of stars. 
The Coulomb logarithm $\ln{\Lambda}=\ln{(R_{\rm{e}}/b_{\rm{min}})}$ 
measures the contribution from stars with different impact parameters 
ranging from $R_{\rm{e}}\simeq 5\, \rm{pc}$,
the effective radius of the NSC \citep[e.g.,][]{gallego-cano_new_2020} 
to $b_{\rm{min}}\approx 2G\mu/v_{\rm{c}}^{2}$, 
the impact parameter below which 
the approximation of weak encounters fails badly \citep{binney_galactic_2011}.

The relaxation rate in Equation~(\ref{equ:deflect}) 
holds for old NSC stars, 
while the internal two-body relaxation between young disk stars is quite different. 
So we multiply Equation~(\ref{equ:deflect}) by two evolutionary modification factors. 
The first factor accounts for 
the growing disk mass and the evolving mass spectrum of the $500\, \rm{Myr}$--population: 
\begin{eqnarray}
  A_{\mathrm{m}}(t_{\rm{lb}})=
  \frac{M_{\rm{d}}(t_{\rm{lb}})}{M_{\rm{d}}(0)}
  \frac{\mu(t_{\rm{lb}})}{\mu_{0}},
\end{eqnarray}
which is normalized to the disk mass at $t_{\rm{lb}}=0$ and $\mu_{0}=1M_{\odot}$
\citep[e.g.,][]{merritt_evolution_2004}. 
The second factor accounts for the enhancement of two-body relaxation due to 
the flattened distribution and coherent rotation of disk stars 
\citep{rybicki_relaxation_1972,sellwood_relaxation_2013}: 
\begin{eqnarray} \label{equ:Ad}
  A_{\rm{d}}=\frac{1}{\sqrt{3}H_{\rm{d}}^{2}} 
  \frac{\ln{(H_{\rm{d}}r/b_{\rm{min}})}}{\ln{(R_{\rm{e}}/b_{\rm{min}})}}. 
\end{eqnarray}
In the above equation, 
the logarithm factor is of order unity, 
and $H_{\rm{d}}$ is the ratio between the scale height and radius of the disk, 
which increases with the random motions of the disk stars 
(measured by their velocity dispersion $\sigma_{\rm{d}}$) as 
\citep{lissauer_growth_1993,heng_longlived_2010} 
\begin{eqnarray} \label{equ:Hd}
  H_{\rm{d}} 
  \approx \frac{\sigma_{d}}{\sqrt{3}v_{\rm{c}}}, 
\end{eqnarray}
assuming that the random motions grow isotropically 
\citep[e.g.,][]{alexander_constraints_2007}. 
Overall, considering that the stellar disk only makes up $\sim 4\%$ of the NSC mass, 
the two-body relaxation rate in Equation~(\ref{equ:deflect}) is modified to 
\begin{align} \label{equ:Ae}
  T_{\rm{r,m}}^{-1} = A_{\rm{e}}T_{\rm{r}}^{-1}, \quad
  % \\ \nonumber
  % = (96\% + 4\%A_{\rm{d}}A_{\rm{m}})T_{\rm{r}}^{-1}, \\
  A_{\rm{e}} = 96\% + 4\%A_{\rm{d}}A_{\rm{m}}
\end{align}

% During the evolution,  and are replaced with their compact remnants,
% 166
%  and the remnants mass are calculated analytically (Hurley et al. 2000; Spera et al. 2015)

With the relations introduced above, 
we track the secular evolution of the disk in time steps of $\Delta t=1\, \rm{Myr}$ 
with an ensemble of mock disk stars. 
We assume that the newly-formed stars follow a Kroupa IMF \citep{kroupa_variation_2001} 
and have velocity dispersions $\sigma_{\rm{d}}=0.1\sqrt{3}v_{\rm{c}}$ ($H_{\rm{d}}=0.1$). 
Starting from $t_{\rm{lb}}=500\, \rm{Myr}$, 
we generate the first set of mock stars of total mass 
$M_{\rm{d,1}}=\alpha_{\rm{sf}}\Delta t$ through Monte Carlo sampling. 
We evaluate the modification factors and relaxation rate 
with Equation~(\ref{equ:deflect}--\ref{equ:Ae}) 
and update the velocity dispersion and the associated $H_{\rm{d}}$ accordingly. 
Some massive stars would evolve off the main sequence and are discarded, 
where the main-sequence lifetime is calculated analytically following 
\cite{hurley_comprehensive_2000}. 
We then move to the next time step by appending another set of mock stars. 
After $t_{\rm{lb}}=150\, \rm{Myr}$, 
no new mock star is appended while the relaxation process goes on. 

\begin{figure*}
  \centering
  \includegraphics[width=0.5\textwidth]{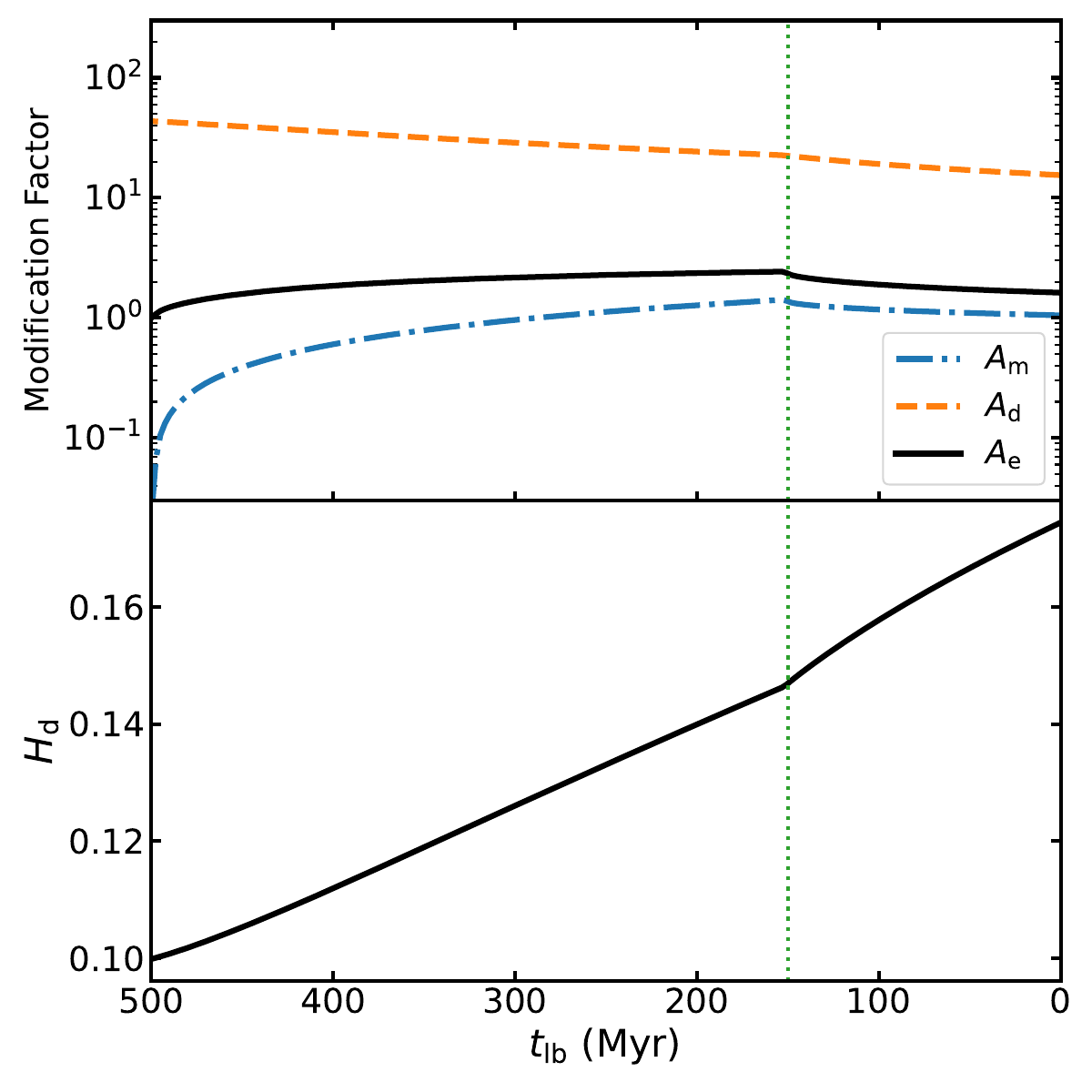}
  \caption{
  Evolution of the modification factors of the relaxation rate (top panel) 
  and the scale height (bottom panel) of the stellar disk,  assuming steady formation of disk stars that 
  ceased at 150$\, \rm{Myr}$ ago (green vertical line).
  The factor $A_{\rm{m}}$ and $A_{\rm{d}}$ accounts for the modification from the 
  evolving stellar population and the disk structure, respectively. 
  The overall modification, denoted as the factor $A_{\rm{e}}$, is 
  a mass-weighted combination of the contribution from the young disk stars and 
  the old NSC stars. 
  % \tg{The star formation is assumed to be steady 
  % and cease at 150$\, \rm{Myr}$ ago (green vertical line).} 
  }  
  \label{fig:Ae}
\end{figure*}

Figure~\ref{fig:Ae} shows the evolution of the modification factors and $H_{\rm{d}}$ 
over $500\, \rm{Myr}$. 
We find that the intra-disk two-body relaxation ($A_{\rm{d}}A_{\rm{m}}$) 
is significantly enhanced.  The enhancement is dominated
% Most of the enhancement is contributed 
by the disk structure ($A_{\rm{d}}=10$--$50$),
which gradually declines as the disk thickens (increasing $H_{\rm{d}}$) with time. 
The factor $A_{\rm{m}}$ basically follows the star formation history, 
while the impact of the mass spectrum is minor. 
Nevertheless, the whole two-body relaxation process ($A_{\rm{e}}$) is 
boosted by only $2$ times due to the limited contribution from the young disk stars 
and remains dynamically insignificant considering its much longer timescale 
\citep[$\sim 10^{10}\, \rm{yr}$;][]{merritt_distribution_2010} than the disk lifetime. 

\section{HVS Production Rate from Disk Stars} \label{sec:rate}

As the stellar disk evolves, 
some disk stars would be ejected as HVSs after close pericenter encounters 
with the SMBH Sgr~A$^{*}$ at $r\leqslant r_{\rm{lc}}$ 
once they fall into a low-angular-momentum region 
in the phase space of ($E$-$L$) called loss cone, bounded by 
\begin{eqnarray} \label{equ:Llc} 
  L^{2}\leq L_{\rm{lc}}^{2}=2r_{\rm{lc}}^{2}[E-\Phi(r_{\rm{lc}})].
\end{eqnarray}
In \cite{cao_recent_2025}, 
we have found that 
the slingshot mechanism by an SMBH-IMBH binary 
of semimajor axis $a_{\rm{MBHB}}\sim 100\, \rm{au}$ 
dominates over the Hills mechanism in ejecting the $500\, \rm{Myr}$--population as HVSs. 
% the late B-type HVSs were predominantly ejected via , 
% while the Hills mechanism contributes a minor fraction. 
This is because during the diffusion toward the SMBH-IMBH binary, 
only stellar binaries of moderate binding energy can produce HVSs, 
while a harder one is more likely to be slingshot, possibly as a hypervelocity binary 
\citep[HVB;][]{lu_hypervelocity_2007,sesana_ejection_2009}, 
and a softer one would end up with the low-velocity ejection of a bound star. 
So here we explore a specific scenario 
where HVSs are slingshot by the SMBH-IMBH binary 
at $r_{\rm{lc}}\simeq a_{\rm{MBHB}}\in [50,\, 400]\, \rm{au}$. 
% where $a_{\rm{MBHB}}$ is the semimajor axis of the SMBH-IMBH binary. 
Nevertheless, our results can be generalized to different $r_{\rm{lc}}$ 
and applied to other close interactions like the HVS ejections in the Hills mechanism. 

To calculate the encounter rate for our disk model, 
we simulate the orbital evolution of disk stars for $500\, \rm{Myr}$ in the GC. 
The first $5\, \rm{Myr}$ of rapid evolution has already been simulated in Section~\ref{subsec:early}. 
However, due to computational resource, 
a full N-body simulation of the subsequent evolution over $495\, \rm{Myr}$ is infeasible. 
Instead, we adopt a mixed numerical--analytic approach. 
% . like that using \texttt{REBOUND} 
% a piecewise strategy: 
We consider that the subsequent evolution is mainly driven by the GC potential 
and perturbed by two-body scatterings from stars of the NSC and NSD, 
while those (gravitational potential and two-body scatterings) 
from the stellar disk are ignored given its insignificant mass fraction 
(see Section~\ref{subsec:secular}). 
In practice, 
for each disk star in the final outcome of the $5\, \rm{Myr}$--simulation, 
its orbit is first integrated in the smoothed GC potential for a radial period.  
Next, we discretize the period into 50 uniform time segments 
and calculate the orbital deflection per segment due to two-body scatterings 
by a fixed population of background stars 
sampled from the NSC and NSD using \texttt{AGAMA} \citep{vasiliev_agama_2019}. 
According to the summed orbital deflection, 
we update the orbit and proceed to the next radial period. 
With the above procedure, we track the evolution of $3\times 10^{6}$ disk stars 
for each disk configuration in Section~\ref{subsec:early}.

\begin{figure*}
  \centering
  \includegraphics[width=0.95\textwidth]{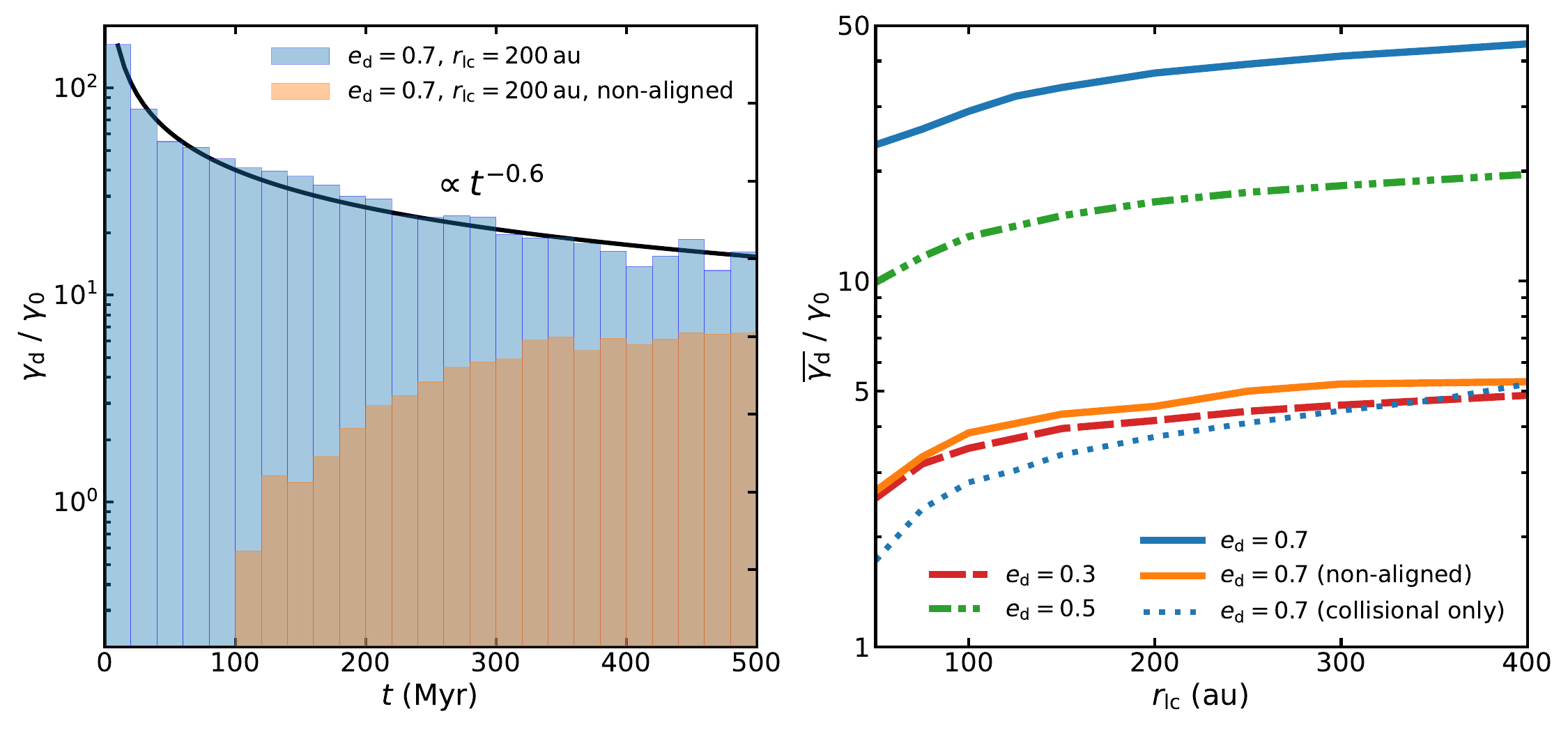}
  \caption{
  Left: 
  Evolution of the close-encounter rate at $r\leq r_{\rm{lc}}= 200\, \rm{au}$ 
  per disk star ($\gamma_{\rm{d}}$), 
  normalized to that for a relaxed population ($\gamma_{\rm{0}}$). 
  The blue (orange) histogram shows 
  the rate for an initially aligned (non-aligned) eccentric disk of $e_{\rm{d}}=0.7$. 
  % the rate peaks at early times and 
  % decays with time approximately following $\propto t^{-0.6}.$ 
  % \tg{By contrast, the rate for a non-aligned eccentric disk of $e_{\rm{d}}=0.7$ (orange) 
  % For clarity, the data have been scaled up by a factor of 3}
  Right: 
  Close-encounter rate per disk star averaged over $500\, \rm{Myr}$ 
  ($\overline{\gamma_{\rm{d}}}$) for different $e_{\rm{d}}$ and $r_{\rm{lc}}$, 
  also normalized to $\gamma_{\rm{0}}$. 
  % Stars are considered to 
  % \tg{form in initially aligned eccentric disks of} $e_{\rm{d}}=0.3, 0.5, 0.7$ 
  % and encounter the SMBH at \tg{$r\leq r_{\rm{lc}}\in [50,\, 400]\, \rm{au}$}. 
  For $e_{\rm{d}}=0.7$, 
  the rate for an initially non-aligned disk is shown as the orange line 
  and that when the stellar orbital evolution is driven solely by two-body relaxation 
  (collisional only) is shown as the dotted blue line. 
  } 
  \label{fig:Plc}
\end{figure*}

The simulation results are presented in Figure~\ref{fig:Plc}. 
The left panel shows the evolution of 
the instantaneous encounter rate per star ($\gamma_{\rm{d}}$) 
over $500\, \rm{Myr}$ post-formation 
when $e_{\rm{d}}=0.7$ and $r_{\rm{lc}}=200\, \rm{au}$. 
For stars forming in an initially aligned disk, 
% where  are assumed to form in an initially aligned disk 
% of eccentricity . 
% like that in Run (C) and (D) in Section~\ref{subsec:early}, 
the rate is significantly enhanced compared to that estimated for a relaxed population: 
$\gamma_{\rm{0}}\simeq 1/[T_{\rm{r}}\ln{(L_{\rm{c}}/L_{\rm{lc}})}]$
\citep{lightman_distribution_1977}. 
% For an initially aligned disk and $r_{\rm{lc}}=200\, \rm{au}$,} 
The rate enhancement ($\gamma_{\rm{d}}/\gamma_{\rm{0}}$) % peaks 
is as large as $\sim 160$ at the beginning and 
diminishes over time approximately as $\propto t^{-0.6}$. 
Consequently, a disk star is very likely to be ejected as a HVS
shortly after its formation ($\leq 50\, \rm{Myr}$), 
which agrees with the observation of the GC-origin HVS1 
\citep[$45\pm 32\, \rm{Myr}$;][]{brown_nature_2012}. 
By contrast, an initially non-aligned disk, 
like in Run (D) in Section~\ref{subsec:early}, 
leads to a much lower rate and 
no encounters within the first $\sim 100\, \rm{Myr}$. 
This is because without the early rapid $L$--evolution due to eccentric disk instability, 
stars have to evolve for a period of time to fill the angular-momentum gap 
between their birthplaces and the loss cone 
\citep[e.g.,][]{merritt_loss_2005,lezhnin_suppression_2015}. 
% their orbits have already experienced the  
% see Section~\ref{subsec:early}
% and been deposited at low-angular-momentum region around the loss cone 
% (see Figure~\ref{fig:rebound}). 
% completely different rate evolution where 
The rate enhancements for different $r_{\rm{lc}}$ and other $e_{\rm{d}}$ evolve similarly 
% have similar evolution trends 
and differ only in magnitude. 
We measure the magnitudes of the rate enhancements with $\overline{\gamma_{\rm{d}}}$, 
the averaged $\gamma_{\rm{d}}$ over $500\, \rm{Myr}$. 
As shown in the right panel, 
the averaged rate enhancement ($\overline{\gamma_{\rm{d}}}/\gamma_{0}$) 
increases mildly with $r_{\rm{lc}}$, 
approximately following $\overline{\gamma_{\rm{d}}}/\gamma_{0} \propto r_{\rm{lc}}^{0.4}$. 
It is the highest for a $e_{\rm{d}}=0.7$ disk 
and about $3(9)$ times lower when $e_{\rm{d}}=0.5(0.3)$.

The enhanced and decaying encounter rate is attributed to two underlying mechanisms: 
(1) the collisionless relaxation due to gravitational torque
in the axisymmetric GC potential 
\citep{magorrian_rates_1999,vasiliev_losscone_2013,penoyre_disruptions_2025}; 
and (2) the radial velocity anisotropy of the disk stars \citep{stone_delay_2018}. 
To quantify their individual contributions for our disk model, 
we conduct controlled experiments for a $e_{\rm{d}}=0.7$ disk where only 
the two-body (collisional) relaxation channel is considered in the simulation. 
We find that when the collisionless relaxation is absent, 
$\overline{\gamma_{\rm{d}}}$ is reduced by a factor of $\sim 10$ 
(dotted line in the right panel of Figure~\ref{fig:Plc}), 
but still about 2--4 times larger than $\gamma_{\rm{0}}$. 
We thus conclude that the collisionless relaxation dominates the rate enhancement, 
while the radial velocity anisotropy plays a subordinate role.

\begin{figure*}
  \centering
  \includegraphics[width=0.95\textwidth]{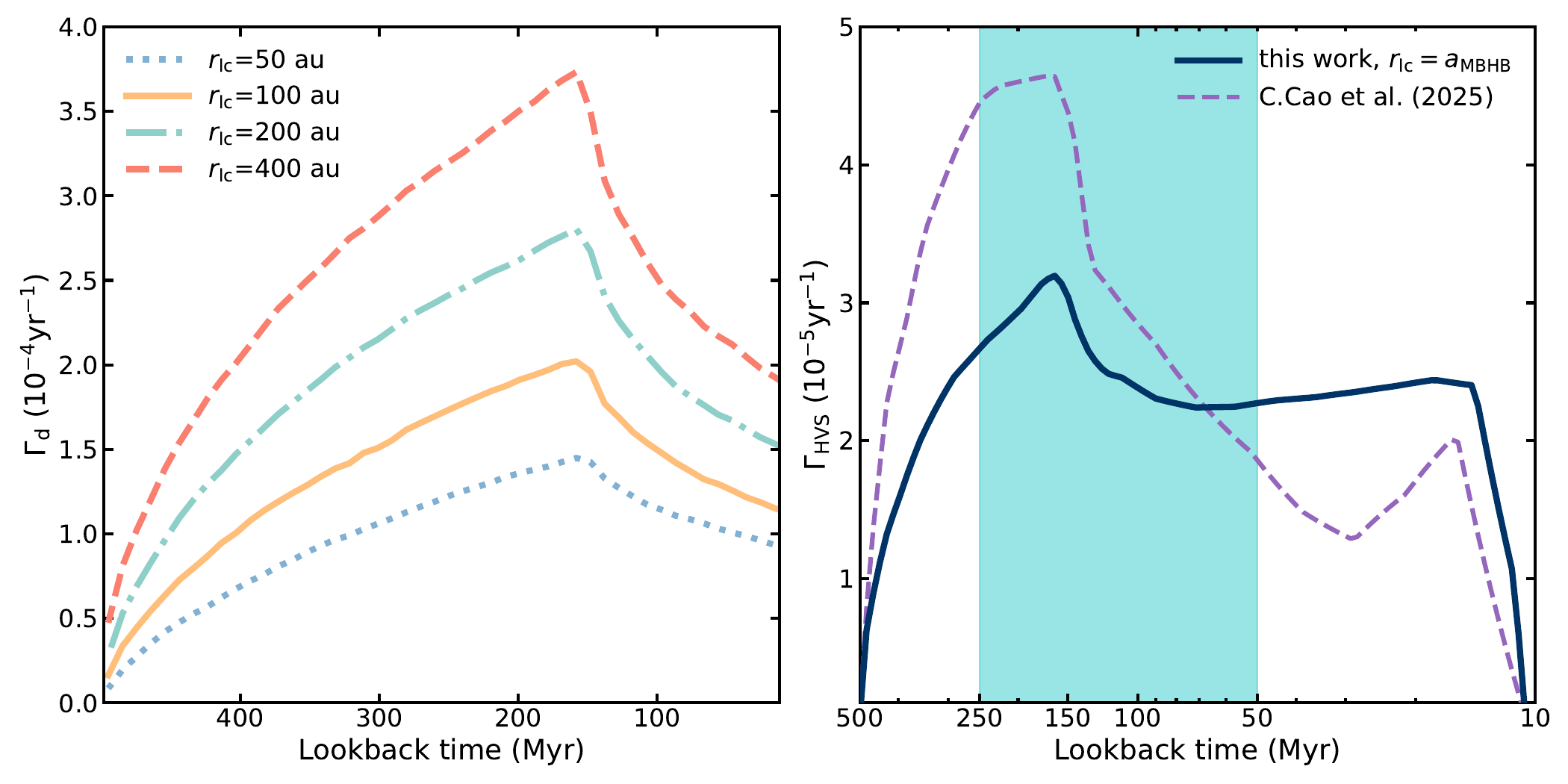}
  \caption{  
  Left: Encounter rate ($\Gamma_{\rm{d}}$) as a function of lookback time 
  for young stars formed  in a $e_{\rm{d}}=0.7$ disk 
  steadily over the past $150$--$500\, \rm{Myr}$. 
  Right: Production rate of HVSs from a $e_{\rm{d}}=0.7$ disk 
  via gravitational slingshot by an SMBH-IMBH binary, 
  which has a mass ratio of $5\times 10^{-3}$ 
  and evolves from a semimajor axis $a_{\rm{MBHB}} \simeq 160\, \rm{au}$ 
  at $500\, \rm{Myr}$ ago to the final coalescence at $\simeq 10\, \rm{Myr}$ ago 
  mainly driven by gravitational wave radiation. 
  The dashed purple curve 
  shows the HVS production rate obtained 
  from the empirical model in \cite{cao_recent_2025}, 
  which assumes an analytical encounter rate 
  that decreases exponentially after $150\, \rm{Myr}$ ago.
  % the star formation. 
  The cyan shadow region indicates the lookback time of $50$--$250\, \rm{Myr}$ 
  when the late B-type HVSs in the halo were ejected.
  }
  \label{fig:rate}
\end{figure*}

Given the above simulation results, 
the encounter rate for our disk model can be computed after convoluting with 
the disk formation history (see Equation~(\ref{equ:disk_formation})) as: 
\begin{eqnarray} \label{equ:Gamma_d}
\Gamma_{\rm{d}}(t_{\rm{lb}})=
\alpha_{\rm{sf}} 
\int^{500\, \rm{Myr}}_{\rm{max}(t_{\rm{lb}},150\, \rm{Myr})}  
\gamma_{\rm{d}}(t=t'-t_{\rm{lb}})\rm{d}t'.
\end{eqnarray}
The left panel of Figure~\ref{fig:rate} shows 
the rate expected for a $e_{\rm{d}}=0.7$ disk. 
The rate is of order $10^{-4}\, \rm{yr}^{-1}$. 
It increases gradually with the stellar formation, 
reaches a peak at $150\, \rm{Myr}$ ago, and declines thereafter. 
And as we demonstrate in Figure \ref{fig:Plc}, 
the rates for different $e_{\rm{d}}$ and $r_{\rm{lc}}$ 
have the same evolution trends and 
differ only by scaling factors of $\overline{\gamma_{\rm{d}}}$. 
For comparison, the traditional loss-cone formalism \citep{lightman_distribution_1977}
estimates a rate that is steady and two orders of magnitude lower: 
\begin{eqnarray} \label{equ:Gamma_0}
  \Gamma_{0}\simeq 4\% \int_{r_{\rm{in}}}^{r_{\rm{out}}} 
  % \frac{1}{T_{\rm{r}}\ln{(L_{\rm{c}}/L_{\rm{lc}})}}
  \gamma_{0}
  \frac{4\pi r^2 \rho}{\langle m \rangle}\mathrm{d}r 
  \simeq 2\times 10^{-6} \rm{yr}^{-1}, 
\end{eqnarray}
assuming that the $500\, \rm{Myr}$--population have completely relaxed and thus 
follow the same density profile ($\rho$) as the NSC.

A fraction ($\eta$) of close encounters would produce HVSs with unbound velocities, 
while others end up with low-velocity ejections of bound stars. 
The HVS production rate ($\Gamma_{\rm{HVS}}$) is thus expressed as  
\begin{eqnarray} \label{equ:Gamma_hvs}
  \Gamma_{\rm{HVS}}=\eta \Gamma_{\rm{d}}. 
\end{eqnarray}
The right panel of Figure~\ref{fig:rate} shows the 
production rate of young HVSs 
slingshot by an SMBH-IMBH binary of mass ratio $5\times 10^{-3}$ 
evolving from $a_{\rm{MBHB}}\simeq 160\, \rm{au}$ at $500\, \rm{Myr}$ ago 
(see Figure 2 in \cite{cao_recent_2025}). 
% Here we explore the production rate of young HVSs of the $500\, \rm{Myr}$--population
% via gravitational slingshot after close encounters with a SMBH-IMBH binary, 
% which is assumed to have a mass ratio of $5\times 10^{-3}$ and 
% a semimajor axis decaying from $\sim 160\, \rm{au}$ at $500\, \rm{Myr}$ ago 
% The result is shown in the right panel of Figure~\ref{fig:rate}. 
The fraction $\eta$ increases with time as the SMBH-IMBH binary decays 
and the slingshot ejections become more energetic \citep[e.g.,][]{yu_ejection_2003}, 
giving rise to a small burst of HVS ejections 
just before the coalescence of the SMBH-IMBH binary. 
Beyond that, the HVS production rate closely follows the close-encounter rate. 
In the past $50$--$250\, \rm{Myr}$ when 
the late B-type HVSs in the halo were ejected (cyan shadow region), 
the averaged fraction is $\eta\approx 10\%$ and 
the typical HVS production rate is $2$--$3\times 10^{-5}\, \rm{yr}^{-1}$, 
which is consistent with the observation \citep{brown_mmt_2014,brown_gaia_2018} 
and the result
\footnote{The rate is calibrated such that 
the model-expected number of late B-type HVSs in the halo is $N_{\rm{HVS}}=33$
instead of $N_{\rm{HVS}}=50$ as used in \cite{cao_recent_2025}, 
because many HVSs in the MMT survey are later found to have non-GC origins 
\citep[e.g.,][]{han_hypervelocity_2025}.}
(dashed purple curve) from the empirical model in \cite{cao_recent_2025} 
at order-of-magnitude level.

\begin{figure*}
  \centering
  \includegraphics[width=0.45\textwidth]{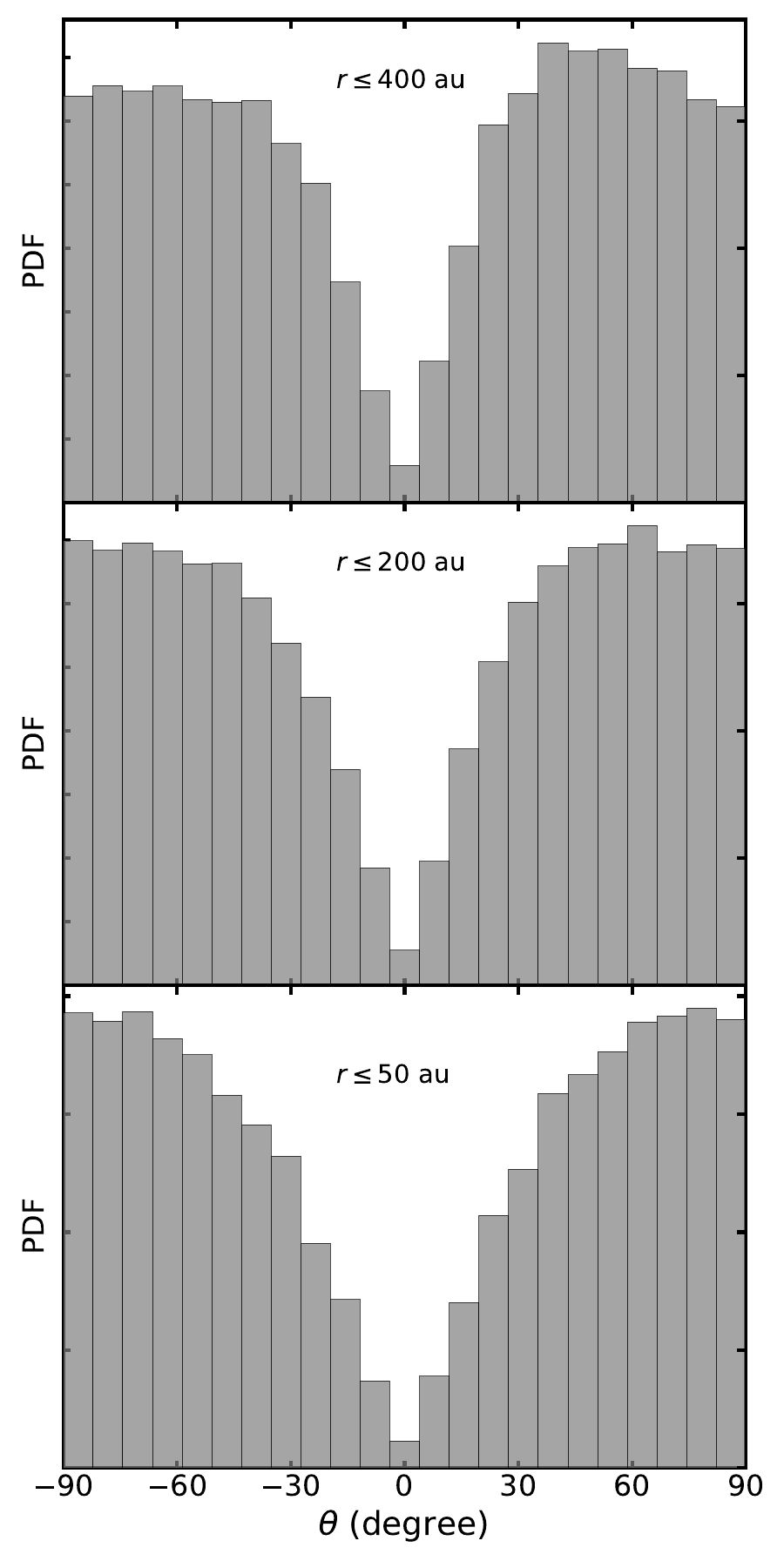}
  \caption{
  The distribution of inclination ($\theta$) of the incident orbits for disk stars 
  that encounter the SMBH at $r\leq 400\, \rm{au}$ (top), 
  $r\leq 200\, \rm{au}$ (middle), and $r\leq 50\, \rm{au}$ (bottom). 
  } 
  \label{fig:angle}
\end{figure*}

Given their initial disk-like structure and evolution in the non-spherical GC potential, 
the $500\, \rm{Myr}$--population are supposed to approach the SMBH on anisotropic orbits 
\citep[e.g.,][]{penoyre_disruptions_2025}. 
Figure~\ref{fig:angle} shows the model-expected inclinations ($\theta$) 
distribution of the incident orbits 
(angles between the angular momentum vectors and the positive z-axis). 
The $\theta$-distribution shows a clear deficit around $\theta\approx 0^{\circ}$ 
and a less obvious peak at 
$\theta\approx (40^{\circ}, 60^{\circ}, 80^{\circ})$ when 
$r_{\rm{lc}}=(400, 200, 50)\, \rm{au}$, 
suggesting that the magnitude and direction of $L$ vary in comparable timescales and 
the injection orbits deviate largely from the original disk plane. 
The large angle deviation is qualitatively consistent with \cite{penoyre_disruptions_2025} 
and warrants attention when mapping the directions of HVSs to their origins 
\citep[e.g.,][]{lu_spatial_2010}.

\section{Conclusions and Discussions} \label{sec:conclusions}

The late B-type HVSs detected in the Galactic halo are suggested to 
have been ejected at a rate 
that significantly exceeds the prediction of the loss-cone theory and 
has declined over the past $\sim 150\, \rm{Myr}$ \citep{cao_recent_2025}. 
This indicates that their progenitors, 
the $500\, \rm{Myr}$--population of the NSC, 
approach the SMBH 
via a mechanism distinct from that of a conventionally relaxed stellar population.
To elucidate this poorly understood process, 
we have constructed an eccentric stellar disk model 
for the young stars of the $500\, \rm{Myr}$--population, 
tracked their orbital evolution in the realistic GC environment, 
and computed the rate at which they are ejected as HVSs.
We found that the migration of the newly formed disk stars toward the SMBH 
are substantially more rapid than that driven solely by two-body relaxation, 
accelerated initially by the eccentric disk instability 
in the aligned eccentric disk (Figure~\ref{fig:rebound}) 
and subsequently by the gravitational torque from the axisymmetric GC potential, 
leading to a strong burst of close encounters within several tens of Myr 
(Figure~\ref{fig:Plc}). 
This, combined with the formation history of the $500\, \rm{Myr}$--population, 
yields a HVS production rate that gradually increases during the star formation 
and quickly drops after the formation ceases at $\sim 150\, \rm{Myr}$ ago. 
For young stars forming in a fairly eccentric disk ($e_{\rm{d}}\gtrsim 0.4$), 
their slingshot interactions with a SMBH-IMBH binary proposed in \cite{cao_recent_2025} 
produce HVS at a rate of
% and ejected by a SMBH-IMBH binary via gravitational slingshot interactions, 
% the production rate can reach 
$10^{-5}$--$10^{-4}\, \rm{yr}^{-1}$, 
which is orders of magnitude larger than that expected for a relaxed population 
\citep{yu_ejection_2003}. 
The enhanced and time-evolving rate is consistent with the observations. 

Our disk model is expected to produce a number of observable signatures in HVSs,
which would be the focus of our next work. 
The most straightforward one is the inhomogeneous radial distribution 
and anisotropic angular distribution of the $500\, \rm{Myr}$--HVSs, 
given their time-evolving production rate (Figure \ref{fig:rate}) and 
highly inclined incident orbits 
relative to the original disk plane (Figure \ref{fig:angle}). 
Another signature comes from 
the preferred ejections of the young $500\, \rm{Myr}$--HVSs over the old ones. 
Such a preference has been previously noticed in observation and 
considered to be evidence for a top-heavy IMF in the GC or 
a HVS ejection scenario that favors massive stars, e.g., the Hills mechanism 
\citep{kollmeier_segue2_2010}. 
Nevertheless, our results suggest that 
the preference could be caused by the boosted supply of young disk stars to the SMBH 
and potentially used to constrain the disk eccentricity (Figure~\ref{fig:Plc}).

For simplicity, this study assumes that young stars of the $500\, \rm{Myr}$--population 
form in a single disk steadily and coherently over the past 500--150$\, \rm{Myr}$ 
(see Equation \ref{equ:disk_formation} and \ref{equ:Gamma_d}). 
However, during the star formation, 
the gas inflow could vary over time, or even temporarily cease for a period, 
leading to a variety of disk assemble history. 
% and the formation and co-evolution of multiple stellar disks in the GC
For example, the NSC stars formed over the past $80\, \rm{Myr}$ \citep{schodel_milky_2020} 
may constitute a distinct disk and co-evolves with the $500\, \rm{Myr}$--disk 
\citep[e.g.,][]{mastrobuono-battisti_star_2019}. 
% As the production rate of young HVSs is directly related to the 
Future HVS observation would provide more specific constraints on the 
dynamics and star formation history in the NSC. 

%%%
\begin{acknowledgments}
This work is supported by the National Natural Science Foundation of China (NSFC No. 11721303), 
China Manned Space Project with No. CMS-CSST-2021-A06, and National Key R\&D Program of China (grant No. 2020YFC2201400). 
X.C. is supported by the National Natural Science Foundation of China (NSFC No. 12473037) and 
S.L. acknowledges the support of the Strategic Priority Research Program of Chinese Academy of Sciences (No. XDB0500203)
and the National Natural Science Foundation of China (NSFC No. 12473017).
\end{acknowledgments}
%%%

\software{
NumPy \citep{harris2020array}, 
SciPy \citep{2020SciPy-NMeth}, 
astropy \citep{2022ApJ...935..167A}, 
pandas \citep{mckinney-proc-scipy-2010}, 
Matplotlib \citep{Hunter:2007}, 
IPython \citep{PER-GRA:2007}
}

% \appendix

% \bibliography{Paper4s}{}
% \bibliographystyle{aasjournalv7}

\end{document}